\documentclass[twocolumn,tighten]{aastex62}

\usepackage{amsmath}
\usepackage{xspace}
\usepackage{multirow}
\usepackage{fancyapj}

\newcommand{\rbr}[1]{\ensuremath{\left( #1 \right)} }

\newcommand{\E}[1]{\ensuremath{\times 10^{#1}} }

\newcommand{\msol}{\ensuremath{M_{\odot}}\xspace}
\newcommand{\ah}{\ensuremath{^{\rm h}}}
\newcommand{\am}{\ensuremath{^{\rm m}}}
\newcommand{\as}{\ensuremath{^{\rm s}}}

\newcommand{\rxte}{\textit{RXTE}\xspace}

\newcommand{\swift}{\textit{Swift}\xspace}

\newcommand{\nicer}{\textit{NICER}\xspace}
\newcommand{\integral}{\textit{INTEGRAL}\xspace}
\newcommand{\src}{Swift~J1756\xspace}

\begin{document}

\title{On the 2018 outburst of the accreting
millisecond X-ray pulsar Swift J1756.9--2508 as seen with NICER}

\author{Peter Bult}
\affiliation{Astrophysics Science Division, 
  NASA's Goddard Space Flight Center, Greenbelt, MD 20771, USA}

\author{Diego Altamirano}
\affiliation{Physics \& Astronomy, University of Southampton, 
  Southampton, Hampshire SO17 1BJ, UK}

\author{Zaven Arzoumanian} 
\affiliation{Astrophysics Science Division, 
  NASA's Goddard Space Flight Center, Greenbelt, MD 20771, USA}

\author{Deepto Chakrabarty}
\affil{MIT Kavli Institute for Astrophysics and Space Research, 
  Massachusetts Institute of Technology, Cambridge, MA 02139, USA}

\author{Keith C. Gendreau} 
\affiliation{Astrophysics Science Division, 
  NASA's Goddard Space Flight Center, Greenbelt, MD 20771, USA}

\author{Sebastien Guillot} 
\affil{CNRS, IRAP, 9 avenue du Colonel Roche, BP
  44346, F-31028 Toulouse Cedex 4, France} 
\affil{Universit\'e de Toulouse, CNES, UPS-OMP, F-31028 Toulouse, France}

\author{Wynn C. G. Ho}
\affil{Department of Physics and Astronomy, Haverford College, 
  370 Lancaster Avenue, Haverford, PA 19041, USA}
\affil{Mathematical Sciences, Physics and Astronomy, and STAG Research Centre,
    University of Southampton, Southampton SO17 1BJ, UK}

\author{Gaurava K. Jaisawal}
\affil{National Space Institute, Technical University of Denmark, 
  Elektrovej 327-328, DK-2800 Lyngby, Denmark}

\author{Steven Lentine}
\affil{Chesapeake Aerospace and Instrument Projects Division,  
  NASA's Goddard Space Flight Center, Greenbelt, MD 20771, USA}

\author{Craig B. Markwardt}
\affiliation{Astrophysics Science Division, 
  NASA's Goddard Space Flight Center, Greenbelt, MD 20771, USA}

\author{Son N. Ngo}
\affil{Mechanical Systems Division, 
  NASA's Goddard Space Flight Center, Greenbelt, MD 20771, USA}

\author{John S. Pope}
\affil{KBRWyle and Software Engineering Division, 
  NASA's Goddard Space Flight Center, Greenbelt, MD 20771, USA}

\author{Paul. S. Ray}
\affiliation{Space Science Division, Naval Research Laboratory,
  Washington, DC 20375-5352, USA}

\author{Maxine R. Saylor}
\affil{KBRWyle and Software Engineering Division, 
  NASA's Goddard Space Flight Center, Greenbelt, MD 20771, USA}

\author{Tod E. Strohmayer} 
\affil{Astrophysics Science Division and Joint Space-Science Institute,
  NASA's Goddard Space Flight Center, Greenbelt, MD 20771, USA}

\begin{abstract}
  We report on the coherent timing analysis of the 182 Hz accreting 
  millisecond X-ray pulsar Swift~J1756.9$-$2508 during its
  2018 outburst as observed with the \textit{Neutron Star 
  Interior Composition Explorer} (\nicer). 
  Combining our \nicer observations with \textit{Rossi X-ray Timing
  Explorer} observations of the 2007 and 2009 outbursts, we also
  studied the long-term spin and orbital evolution of this source. We find that
  the binary system is well described by a constant orbital period model,
  with an upper limit on the orbital period derivative of $|\dot P_b|
  < 7.4\E{-13}$~s~s$^{-1}$. Additionally, we improve upon the
  source coordinates through astrometric analysis of the pulse
  arrival times, finding R.A. = $17\ah56\am57.18\as\pm0.08\as$ 
  and Decl. = $-25\arcdeg06\arcmin27.8\arcsec\pm3.5\arcsec$,
  while simultaneously measuring the long-term spin frequency 
  derivative as $\dot\nu = -7.3\E{-16}$~Hz~s$^{-1}$.
  We briefly discuss the implications of these measurements in
  the context of the wider population of accreting millisecond
  pulsars. 
\end{abstract}


\section{Introduction} \label{sec:intro}
	Accreting millisecond X-ray pulsars (AMXPs, \citealt{Wijnands1998a})
    are rapidly rotating neutron stars whose spin periods can be
    observed directly as coherent oscillations in their X-ray light
    curves. By monitoring the frequency evolution of such pulsations
    on years-long timescales, we can measure the rate of spin change
    and gain insight into intrinsic properties of the 
    neutron star. 
    Additionally, precise timing measurements enable detailed
    studies of the binary orbit, offering potential insight into
    the binary evolution of millisecond pulsars.
    
    Such long-term monitoring efforts, however, are complicated by 
    the transient nature of these AMXPs. The pulsar is visible only 
    during X-ray outbursts, when the source is actively accreting.
    Such outbursts typically last for a number of days to weeks, and
    may be interspersed by years or even decades of quiescence. 

    Of the population of AMXPs currently known \citep{Patruno2012b,
    Strohmayer2017, Sanna2017a, Sanna2018}, only eight sources have shown
    recurrent outbursts, and of those only three could be studied with
    sufficient precision to allow for the long-term spin frequency 
    derivative to be measured.
    Studies considering the binary orbit of AMXPs face a similar
    situation: a physically interesting sensitivity to the orbital period rate of
    change has been achieved for only three AMXPs. One of these
    seems consistent with a slow evolution driven by angular
    momentum loss through gravitational radiation
    \citep{Patruno2017a, Sanna2017b}, the other two evolve on a
    markedly faster timescale (\citealt{Patruno2012a, Sanna2017c},
    although see also \citealt{Patruno2017b} for a detailed
    discussion of these and other classes of binary systems).
    It is therefore of considerable interest to increase the sample size of 
    this population. 
    
    The AMXP Swift~J1756.4$-$2508 (hereafter \src) was first discovered in 
    2007 June \citep{Krimm2007a} and was quickly found to be
    a 182 Hz pulsar \citep{Markwardt2007a}. It was observed in 
    outburst again in 2009 July \citep{Patruno2009e}, but remained in 
    quiescence for the following 9 years. The \textit{Rossi X-ray
    Timing Explorer} (\rxte) observed both outbursts extensively;
    nonetheless, a detailed analysis of those data \citep{Patruno2010c} 
    did not detect a spin frequency change: with only two reported
    outbursts, both of short duration, the upper limit on the
    spin frequency derivative was $|\dot \nu| < 3\E{-13}$~Hz~s$^{-1}$, which 
    is much larger than any neutron star spin-down observed in similar
    sources. 

    On 2018 April 3 \integral reported a new outburst from \src
    \citep{AtelMereminskiy18}.
    Follow-up observations with the Neutron Star Interior Composition
    Explorer (\nicer, \citealt{Gendreau2017}) quickly revealed the
    presence of 182 Hz pulsations \citep{AtelBult18a}, confirming
    the third known outburst of this AMXP. With an observational
    baseline that spans over a decade, we may now probe a
    physically interesting regime of spin-down parameters. In this
    work, we present a coherent timing analysis of
    the \nicer campaign for the recent \src outburst, together
    with archival data for the earlier outburst episodes.

\section{Observations} \label{sec:obs}
    The \nicer X-ray Timing Instrument consists of an array of 56
    concentrator X-ray optics paired with silicon drift detectors
    \citep{Gendreau2016}. 
    These detectors are sensitive in the $0.2-12$ keV energy band
    \citep{Prigozhin2012}, with an energy resolution of better than
    $150$~eV, and a timing precision of $\sim100$~ns rms. We observed
    \src with 52 operating detectors, giving a total effective
    area of $\sim1900$~cm$^2$ at 1.5~keV.
    
    We monitored \src from 2018 Apr 4 until 2018 Apr 25, at which time
    the source had returned to quiescence \citep{AtelBult18b}. For
    this paper we analyzed all available \nicer data (ObsID 1050230101
    through 1050230108), which together amounted to $54$~ks of
    unfiltered exposure.
    
    We processed the data using \textsc{heasoft} version 6.24 and
    \textsc{nicerdas} version 2018-04-06\_V004. The data were cleaned
    using standard filtering criteria: we selected only those epochs
    that had a pointing offset $<54\arcsec$, bright Earth limb angle
    $>40\arcdeg$, dark Earth limb angle $>30\arcdeg$, and were outside
    the South Atlantic Anomaly (SAA). After processing we were left
    with $42$ ks of exposure.

    Next we computed the $12-15$ keV light curve using 8 second bins.
    Above $12$ keV the performance of the detectors and X-ray optics
    has diminished such that essentially no astrophysical signal is
    expected. Nonetheless, we observed several epochs during which
    the count rate in this light curve was greater than 1~ct/s, 
    which we attributed to periods of increased background. Since
    these intervals were correlated with an elevated count-rate in the
    $0.4-10$ keV light curve, we removed them from the analysis. An
    additional $1$~ks of exposure was removed in this way.

    Finally, we were left with $41$~ks of good time exposure. We
    applied barycentric corrections to those events using the
    \textsc{ftool barycorr} with the source coordinates of
    \citet{Krimm2007b} and the DE405 Solar System ephemeris. 
    No X-ray bursts were observed.

    Because \nicer does not have imaging capabilities, we used \nicer
    observations of the \rxte blank field region 8 \citep{Jahoda2006}
    to estimate the background count rate.  Applying the same
    filtering criteria, we obtained $74$ ks of good background field
    exposure, yielding an averaged background rate of $2$~ct/s in the
    $0.4-10$ keV band.

\section{Analysis \& Results} 
\label{sec:results}
    For the coherent timing analysis, we selected all events
    in the $0.4-10$ keV energy range and corrected
    the photon arrival times for the source binary motion
    using the ephemeris reported by \citet{Patruno2010c}.
    Assuming a constant orbital period, we could extrapolate
    the orbital phase to the current epoch, yielding
    a predicted time of passage through ascending node, $T_{\rm asc}$, 
    in terms of MJD(TDB) of
    \begin{equation*}
        T_{\rm asc, pred} =  58211.0170 \pm 0.0002
    \end{equation*}
    As the $1\sigma$ uncertainty on this extrapolation is only about $0.5\%$
    of the orbital period, the orbital solution allowed for a coherent
    propagation across the nine years of quiescence since the last
    outburst.
    To test this prediction, we searched a grid of 
    $T_{\rm asc}$ values in steps of $1\E{-5}$~d spanning one full 
    orbit and folded the data on each trial ephemeris. The 
    highest pulse amplitude was found at $T_{\rm asc, grid} 
    = 58211.01736$, which is consistent with the extrapolated 
    solution. We adopted the results from our grid search as
    our initial trial ephemeris.

    Next, we divided the data into segments of $\sim1000$~s
    exposure. For each segment we used the trial ephemeris to
    remove the orbital modulation and then folded on the
    pulse period. The resulting pulse profiles were fit
    with a constant plus two sinusoids, where one sinusoid was
    set at the spin frequency and the second at twice that
    frequency, so to capture the fundamental and second harmonic,
    respectively. A pulse harmonic was considered to be significant
    when its amplitude divided by its statistical uncertainty
    was greater than three, that is, when $A / \sigma_A > 3$.
    Under this condition a third harmonic was never required.
    Pulse amplitudes are reported in terms of fractional
    rms
    \begin{equation}
        r_i = \frac{1}{\sqrt{2}}\frac{A_i}{N_\gamma - B},
    \end{equation}
    where $A_i$ is the measured sinusoidal amplitude of the $i$th
    harmonic, $N_\gamma$ the total number of photons in the considered
    segment, and $B$ the estimated number of background events in
    that segment. We further note that rms amplitudes are smaller 
    than sinusoidal amplitudes by a factor of $\sqrt{2}$.
    
    As the pulse profiles of AMXPs may change over time
    \citep[see, e.g.][for a review]{Patruno2012b}, we modeled the measured pulse 
    arrival times for each harmonic
    separately. For both harmonics we adopted a timing model
    consisting of a circular orbit and constant spin frequency.
    Hence, our model consisted of four parameters: the binary orbital
    period $P_b$, the projected semi-major axis $a_x \sin i$, the
    time of ascending node $T_{\rm asc}$, and the spin frequency
    $\nu$. We fit this model to the data using \textsc{tempo2}
    \citep{Hobbs2006} and iterated the procedure of folding and 
    refitting until the timing solution had converged.

    The phases of the fundamental pulsation are well
    described by the timing model, with goodness-of-fit statistic
    of $\chi^2=16.6$ for 16 degrees of freedom (dof, see Table 
    \ref{tab:ephemeris}). 
    The second harmonic, on the other hand, shows significant residual
    deviations ($\chi^2=25.7$, 7 dof). These residuals can be attributed 
    to timing noise, which was also present in the second harmonic of 
    the previous outbursts \citep{Patruno2010c}.   

\begin{table}[t]
    \newcommand{\mc}[1]{\multicolumn2c{#1}}
    \caption{%
        Timing solution for the 2018 outburst of \src. 
        \label{tab:ephemeris}
    }
    \begin{center}
    \begin{tabular}{l l l}
        \tableline
        Parameter & {Value} & {Uncertainty} \\
        \tableline
        $\nu$ (Hz)            & 182.0658037800 & 4.5\E{-8} \\
        $a_x \sin i$ (lt-ms)  & 5.981          & 4.6\E{-2} \\
        $P_{b}$ (s)           & 3282.463       & 9.5\E{-2} \\
        $T_{\rm asc}$ (MJD)   & 58211.017496   & 8.4\E{-5} \\
        $\epsilon$            & $<1\E{-2}$     & \\ 
        \tableline
        $\chi^2$/dof          & 16.6 / 16      & ~ \\  
        \tableline
    \end{tabular}
    \end{center}
    \tablecomments{%
        Uncertainties give the $1\sigma$ statistical errors
        and the upper limit is quoted at the 95\% c.l. The $\epsilon$ 
        parameter gives the orbital eccentricity.
    }
\end{table}

\begin{figure}[th]
	\centering
    \includegraphics[width=\linewidth]{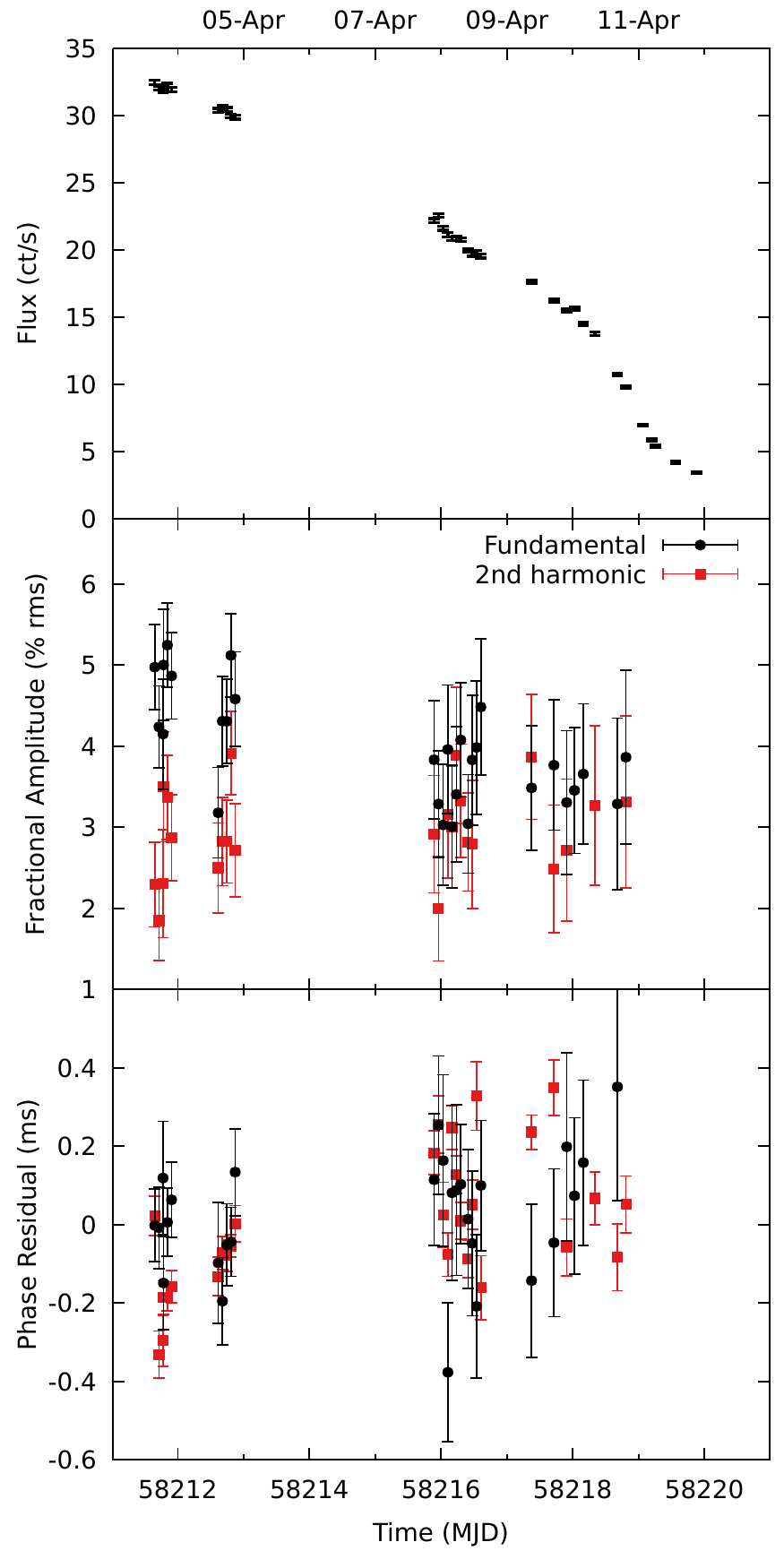}
    \caption{%
        Outburst and pulse evolution of \src for $\sim1$~ks
        segments. 
        Top: $0.4-10$ keV light curve (source+background).
        Middle: fractional amplitudes of the fundamental
        pulsation (black) and second harmonic (red).
        Bottom: phase residuals of the pulsations with
        respect to the ephemeris reported in Table 
        \ref{tab:ephemeris}. Upper limits on non-detections 
        of individual harmonics are not shown.
    }
    \label{fig:lightcurve}
\end{figure}

    The pulse evolution for our best fit timing solution is 
    shown in Figure \ref{fig:lightcurve}. The top panel gives 
    the count-rate in each 
    segment, and shows how the source count-rate decayed steadily
    from April 3 (MJD 58211) to April 12 (MJD 58220). The additional 
    observation on April 25 (MJD 58233) is not shown as the source was 
    in quiescence. The middle panel gives the fractional amplitudes of
    the fundamental pulsation and second harmonic.  Finally, the
    bottom panel gives the residual phase variations.  These residuals
    show that the data are well described by a circular orbit model,
    and no anomalous phase jumps are observed.  The orbital parameters
    of our best-fit solution are shown in Table \ref{tab:ephemeris}.

\begin{figure}
    \centering
    \includegraphics[width=\linewidth]{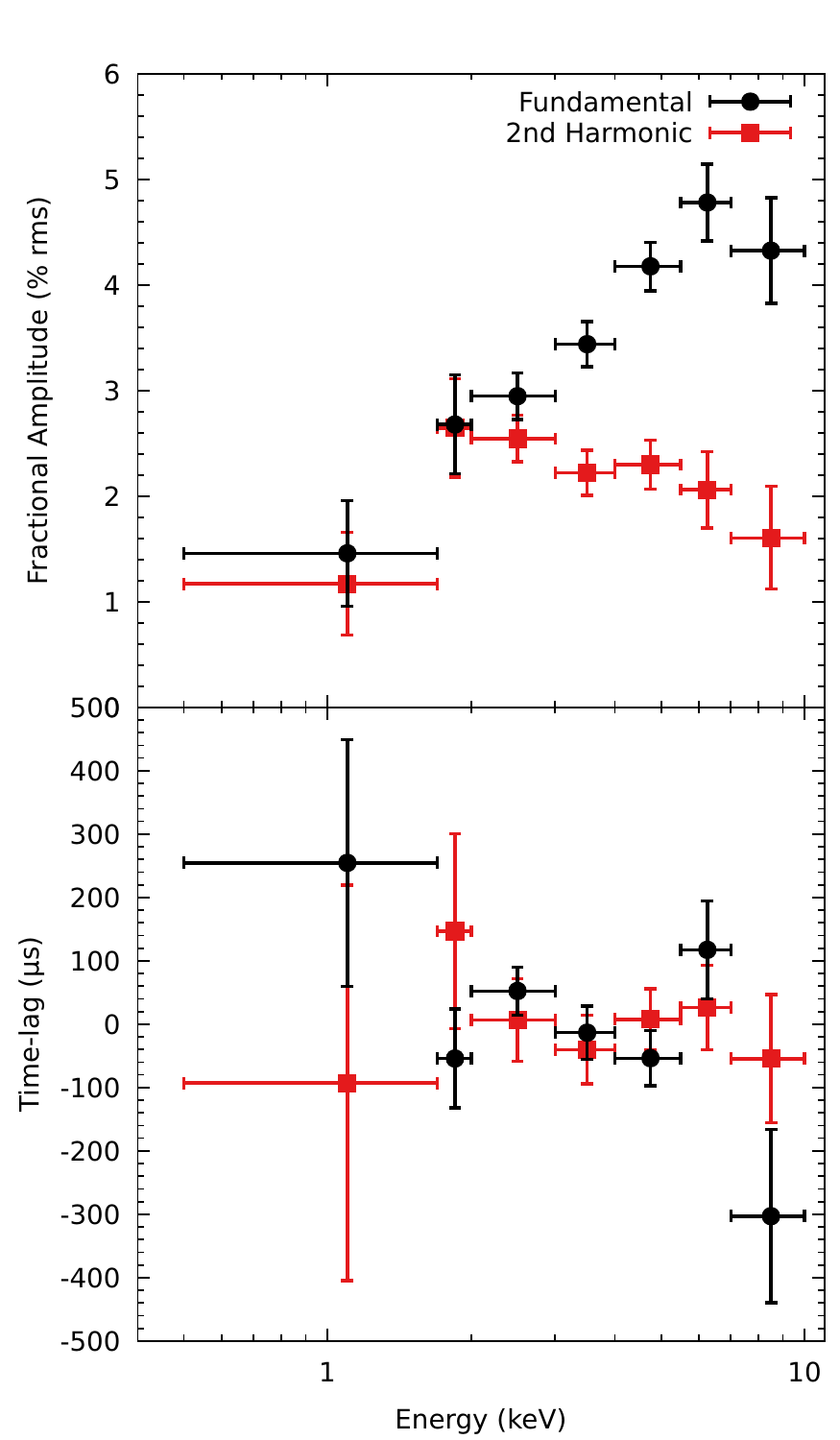}
    \caption{%
        Energy dependent properties for the fundamental (black)
        and second harmonic (red) of the pulsations, with;
        top: pulse fractional amplitude; and 
        bottom: pulse time-lags with respect to the timing 
        model reported in Table
        \ref{tab:ephemeris}.
        \label{fig:pulse.energy}
    }
\end{figure}

    We also considered the energy dependence of the pulsations. We
    divided the $0.4-10$ keV energy range into 7 bins, each about 1
    keV wide. For each bin we then applied the timing solution
    reported in Table \ref{tab:ephemeris} and folded all data 
    into a single pulse profile. We measured the amplitudes of 
    the fundamental and second harmonic, as well as their relative
    phases. As shown in Figure \ref{fig:pulse.energy} (top panel), the
    fractional amplitude of the fundamental increases with
    energy, while the second harmonic shows a slight decline 
    in its fractional amplitude. For both harmonics the phase
    residuals are approximately constant across the \nicer 
    passband (Figure \ref{fig:pulse.energy}, bottom panel). 
    
\subsection{Orbital evolution}
    The orbital period measured for the 2018 outburst of
    \src is consistent with the orbital period reported
    by \citet{Patruno2010c} within their combined $1\sigma$ statistical 
    uncertainty. To obtain a more accurate measure of the 
    orbital evolution, we performed a coherent analysis 
    of the orbital phase across all three outbursts. 
    
    Following the procedure outlined in \citet{Hartman2008}, we 
    compared the predicted evolution of the best known orbital
    ephemeris with the orbital phases measured in each of the
    three outbursts. Specifically, we calculate the residual 
    time of passage through the ascending node, 
    $\Delta T_{\rm asc}$, as
    \begin{equation} \label{eq:tasc}
        \Delta T_{\rm asc} = T_{{\rm asc}, i} - (T_{\rm ref} + N P_b),
    \end{equation}
    where $T_{{\rm asc}, i}$ is the time of ascending node
    for the $i$th outburst, $N$ is the integer number of orbital
    cycles between the $i$th outburst and the reference time,
    and we used the reference time, $T_{\rm ref}$, and
    orbital period $P_b$ as reported in Table 4 of \citet{Patruno2010c}.
    
    \begin{figure}[t]
        \centering
        \includegraphics[width=\linewidth]{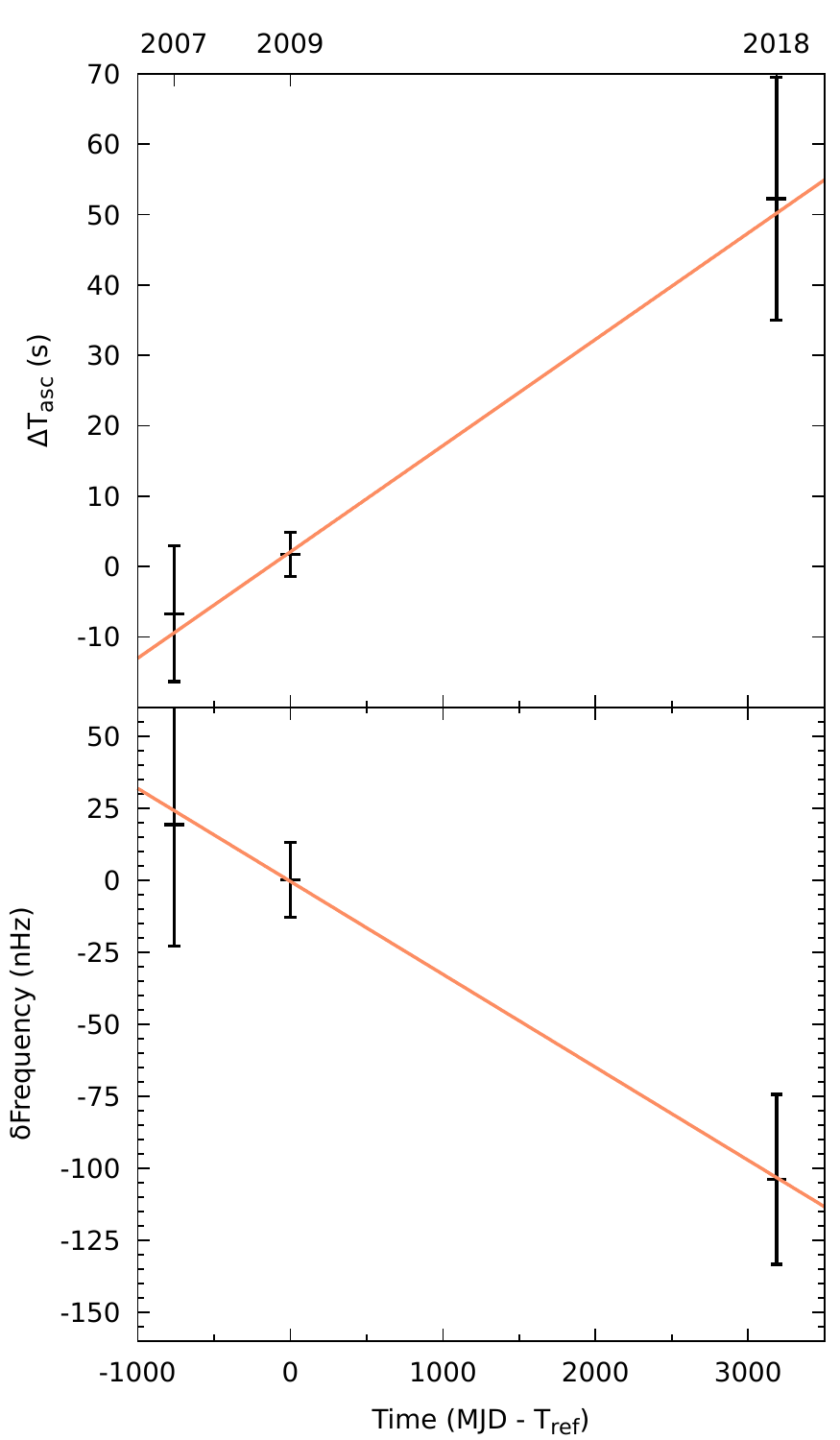}
        \caption{
            Long term evolution of \src, showing;
            top: evolution of the time of ascending node 
            with respect to the ephemeris of \citet{Patruno2010c};
            and
            bottom: pulse frequencies relative to offset frequency 
            $\nu_0~=~182.065803903$ Hz for each of the three 
            outbursts. 
            The reference time is $T_{\rm ref} = 55026.03429$.
            Solid lines show best-fit models (see text for details). 
        }
        \label{fig:orbit}
    \end{figure}

    As shown in Figure \ref{fig:orbit} (top panel), we found that 
    the residual $T_{\rm asc}$ shows a steady advance over time,
    indicating that the orbital period, $P_{b, {\rm trial}}$, used to
    obtain these values underestimates the actual period.  Indeed,
    these residuals are poorly described by a constant
    ($\chi^2 = 9.3$, 2 dof), and instead prefer a linear model as 
    \begin{equation}
        \Delta T_{\rm asc} = N \delta P_b,
    \end{equation}
    with a best-fit statistic of $\chi^2 = 0.6$ for 2 degrees of
    freedom.  This fit gives us a correction to (improvement of) the
    constant orbital period of $\delta P_b = 5.8\E{-4}$~s, such
    that $P_b = P_{b, \rm trial} + \delta P_b$; however, the very low
    $\chi^2$ suggests that the uncertainties obtained from this fit
    may not be reliable.
    
    For a more robust estimate of the long term orbital evolution, 
    we instead analyze all three outbursts simultaneously. We first
    reconstructed the pulse arrival times of the 2007 and 2009
    outbursts, by repeating the analysis procedures described in
    \citet{Patruno2010c}. We then fit our timing model to all 
    three outbursts at once. In this fit the orbital parameters were
    coupled, and the spin frequency was left free for each of 
    the three outbursts \citep[see, e.g.,][]{Bult2015c}.
    This procedure gives a good fit to the data ($\chi^2=226.2$,
    197 dof), and yields an orbital period correction of 
    $\delta P_b = (5.2\pm0.5)\E{-4}$~s, which is consistent with the previously
    mentioned linear fit. The complete set of best-fit orbital 
    parameters is reported in Table \ref{tab:best.ephemeris}. 
    
    An orbital period derivative was not required to obtain a good
    fit to outbursts of \src, hence we found no evidence that
    the orbital period changed over the observed time-span of 11
    years. By adding this parameter to the joint-fit procedure, we
    obtained a $95\%$ confidence level upper limit on the orbital
    period derivative of $|\dot P_b| < 7.4\E{-13} \mbox{~s s}^{-1}$. 

\subsection{Spin frequency evolution}
\label{sec:f.evolution}
    The joint analysis described in the previous section gave 
    us a local spin frequency measurement for each of the three
    outbursts. The measured frequencies for the 2007 and 2009
    outbursts (Figure \ref{fig:orbit}; bottom panel) were consistent with 
    those reported by \citep{Patruno2010c} within their $1\sigma$ 
    statistical uncertainties. Combined with the spin frequency measured
    for the 2018 outburst as observed with \nicer, we found a
    clear decline in spin frequency over time. Indeed, a
    constant spin frequency model gave a poor description of these
    data ($\chi^2 = 11.1$, 2 dof), whereas a linear model of the form
    \begin{equation}
        \Delta \nu = \delta\nu + \dot\nu T
    \end{equation}
    did better ($\chi^2=0.02$ for 1 degrees of freedom). The 
    spin frequency derivative implied by this fit is on the order 
    of $-4\E{-16}$~Hz~s$^{-1}$. These measurements, however,
    are subject to a systematic bias associated with the uncertainty
    of the source coordinates \citep{Manchester1972}. The best available
    source coordinates were obtained with \textit{Swift}/XRT 
    \citep{Krimm2007b} and have a comparatively large uncertainty
    of $3.5\arcsec$ (90\% c.l.). A first-order estimate of the effect that this 
    uncertainty has on the spin frequency derivative 
    \citep{Burderi2007,Hartman2008} gives $\sigma_{\dot\nu, 
    {\rm pos}} \sim 10^{-15}$ Hz~s$^{-1}$, which is comparable to the 
    slope observed in Figure \ref{fig:orbit}. Hence, a more careful 
    analysis is required.

\begin{figure*}
    \centering
    \includegraphics[width=\linewidth]{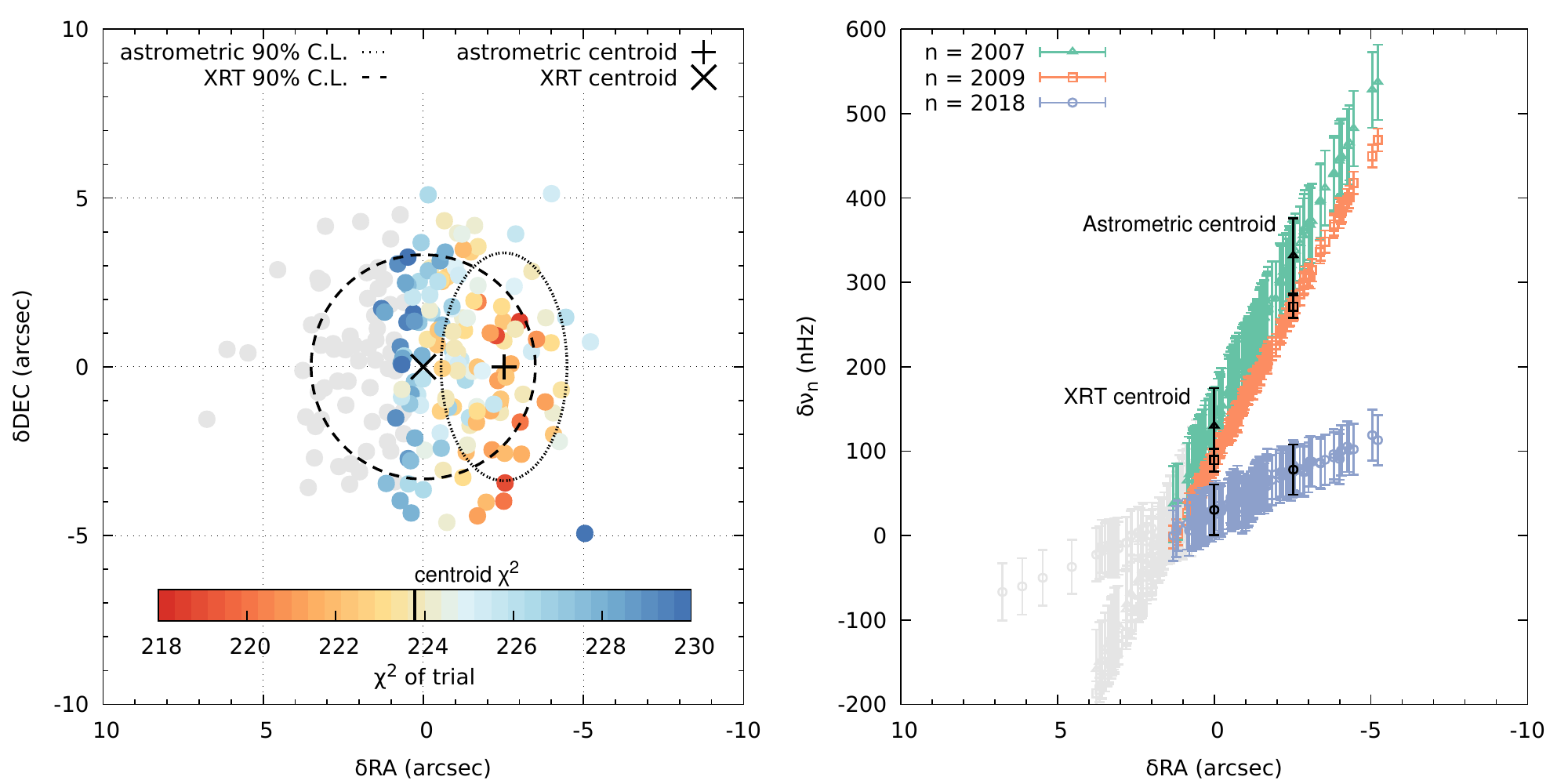}
    \caption{%
        Results from a Monte Carlo (MC) study of the influence of the
        source position uncertainty on the performance of the
        timing model. 
        Left: the MC trials relative to the centroid of the \swift/XRT
        source position \citep{Krimm2007b}, with the color coding
        indicating the $\chi^2$ of the best-fit timing model for 
        those coordinates. Also shown are the \swift/XRT centroid
        position and the astrometric position derived in this work,
        along with their respective 90\% c.l. contours.
        Right: the spin frequency measured per outburst for the MC
        trials. Spin frequencies are shown relative to $\nu_0$ (see Figure 
        \ref{fig:orbit}). The black points mark the spin frequencies retrieved 
        at the astrometric and \swift/XRT centroid positions, as
        indicated. In both panels points in gray represent MC trials that were
        rejected by the timing model fit (see section \ref{sec:f.evolution} 
        for details).  
        \label{fig:mc}
    }
\end{figure*}

    To assess the effects of the source position uncertainty on
    our timing analysis, we generated 500 random coordinates 
    distributed according to the \swift/XRT error circle. For each
    trial position we reapplied the barycentric corrections, and fit the
    timing model to the three outbursts jointly. We then measured,
    as a function of $\delta$RA and $\delta$DEC relative to the 
    \swift/XRT centroid, the $\chi^2$ of the timing model fit (which
    has 196 dof), and the spin frequency in each of the three
    outbursts. 
    The results of these fits are summarized in Figure \ref{fig:mc}.

    Shown in the left panel are the $\chi^2$ values of the timing
    model fit, with the colored points indicating the trials for which
    this fit was statistically acceptable (p-value better than 0.05).
    Clearly the timing model is sensitive to $\delta \mbox{RA}$,
    implying we can refine the source position. The timing model
    cannot constrain $\delta \mbox{DEC}$, however, given that \src 
    is located only $1.67\arcdeg$ away from the ecliptic, this is 
    not surprising. 
    As shown in the right panel of Figure \ref{fig:mc}, we 
    additionally found that the spin frequencies measured
    per outburst tend to diverge for decreasing $\delta$RA, 
    which is the region of parameter space that is clearly favored 
    by the timing solution. 
    To capture both effects, we searched for the minimum of
    the $\chi^2$ space as a function of $\delta$RA\footnote{%
        Note that this is equivalent to including RA as a free
        parameter in the timing model.
    } and scanned the $\chi^2$ space out to $\Delta\chi^2=1$ to determine
    the position uncertainty. This gave a best-fit position of 
    $\delta \mbox{RA} = -2.5 \pm 1.1\arcsec$.     
    We then measured the spin frequency derivative at the contours of our
    scan in $\delta$RA to determine the range of allowed values.
    Adding also (in quadrature) the statistical uncertainty of the
    linear fit to the long term spin frequency trend, we then arrive at a
    spin frequency derivative measurement of $\dot\nu =
    (-7.3\pm2.6)\E{-16}$~Hz~s$^{-1}$ (see Table
    \ref{tab:best.ephemeris} for the complete best-fit timing
    solution, including the refined source position).

\begin{table}[t]
    \newcommand{\mc}[1]{\multicolumn2c{#1}}
    \caption{%
        Best-fit timing parameters of \src from the 
        joint analysis of the 2007--2018 outbursts.
        \label{tab:best.ephemeris}
    }
    \begin{center}
    \begin{tabular}{l l l}
        \tableline
        Parameter & {Value} & {Uncertainty} \\
        \tableline
        R.A. (J2000)                & 17\ah56\am57.18\as& 0.08\as \\
        Decl.(J2000)               &$-25\arcdeg06\arcmin27.8\arcsec$& 3.5\arcsec \\
        \tableline
        $\nu_0$ (Hz)                & 182.065804074     & 8.3\E{-8}  \\
        $\dot\nu$ (Hz/s)            &$-7.3\E{-16}$      & 2.6\E{-16} \\
        $P_{b}$ (s)                 & 3282.352018       & 4.7\E{-5}  \\
        $|\dot P_{b}|$ (s s$^{-1}$) & $<7.4\E{-13}$     & ~          \\
        $a_x \sin i$ (lt-ms)        & 5.965             & 1.3\E{-2}  \\
        $T_{\rm asc}$ (MJD)         & 55026.034350      & 1.4\E{-5}  \\
        $\epsilon$                  & $<1\E{-2}$        & ~          \\ 
        \tableline
        $\chi^2$/dof                & 218.5 / 196       & ~ \\  
        \tableline
    \end{tabular}
    \end{center}
    \tablecomments{%
        The spin frequency reference epoch is set at MJD $55026.6$.
        Uncertainties give the $1\sigma$ statistical error and upper
        limits are quoted at the 95\% c.l., with $\epsilon$ giving the
        binary eccentricity. Declination was not included in the
        fit. 
    }
\end{table}

\section{Discussion}
    We reported on the coherent timing analysis of the 2018
    outburst of \src as observed with \nicer. Consistent with
    analyses of the previous outbursts \citep{Krimm2007b,
    Patruno2010c}, we find that the X-ray pulsations have energy 
    dependent amplitudes; the fractional amplitude of the fundamental
    increases with energy, whereas the fractional amplitude of
    the harmonic shows a slight decline with energy. This
    energy dependent behavior is not unusual in AMXPs
    \citep{Patruno2012b} and can be interpreted in terms
    of the thermal emission from the stellar hotspot and reprocessing
    in the accretion column \citep[e.g.][]{Gierlinski2002, Ibragimov2009}.

    The pulse arrival times of the 2018 outburst are well described
    by a timing model consisting of a circular orbit with a constant 
    spin frequency. The pulse phases with respect to this model do not
    show spurious residuals with time or orbital phase, and no
    evidence is found that the pulse arrival times exhibit an
    additional delay associated with passing through the gravitational
    well of the companion star (Shapiro delay). We note, however, that 
    the expected Shapiro delay is given as \citep{Shapiro1971}
    \begin{equation}
        \Delta t_S(\Phi) = -2 \frac{GM_C}{c^3} \rbr{1 - \sin i \sin \Phi },
    \end{equation}
    with $\Phi$ the orbital phase, $G$ the gravitational constant,
    $c$ the speed of light, and $i$ the inclination. Even for the maximum allowed companion 
    mass, $M_C=0.030\msol$ \citep[but see section \ref{sec:orb} for
    more details]{Krimm2007b} and an inclination of 90\arcdeg, the 
    largest delay we can expect is only $4 \mu s$. As this time-delay
    is smaller than the uncertainty on our phase residuals by nearly
    two orders of magnitude (see Figure \ref{fig:lightcurve}), we are
    not sensitive to Shapiro delays in \src.

    Comparing our measurements for the 2018 outburst with those 
    of the 2007  and 2009 outbursts as observed with \rxte, we 
    analysed the long term evolution of this source. We found that 
    the binary system is consistent with having a constant 
    orbital period and that the pulsar shows a spin frequency 
    derivative of $\dot\nu = -7.3\E{-16} \mbox{~Hz s}^{-1}$. 

\subsection{Spin-down evolution}
    The long term spin frequency derivative measured in \src is of the
    same order as the spin frequency derivatives measured in other
    AMXPs \citep{Hartman2008, Patruno2010e, Riggio2011}.  Most likely,
    this frequency change is driven by the neutron star's loss of
    rotational energy. If so, then the spin-down luminosity is given
    as
    \begin{align}
        \dot E_{\rm sd} 
          &= 5 \E{33}~\rbr{\frac{I}{10^{45}~\mbox{g cm}^{2} }} 
          \rbr{\frac{\nu}{182~\mbox{Hz}}} \nonumber \\
          &\times \rbr{\frac{-\dot\nu}{7.3\E{-16}~\mbox{Hz s}^{-1}}}
        \mbox{erg s}^{-1},
    \end{align}
    where $I$ represents the neutron star moment of inertia. 

    The long term spin-down of a neutron star is usually assumed
    to be dominated by the braking torque associated with a spinning
    magnetic field. Assuming this mechanism is responsible for the
    observed spin-down in \src, we can compute the magnetic dipole 
    moment as \citep{Spitkovsky2006}
    \begin{align}
        \mu &= 2.9\E{26}~\rbr{1 + \sin^2\alpha}^{-1/2} \nonumber \\
            &\times \rbr{\frac{I}{10^{45}\mbox{~g cm}^2}}^{1/2} 
                    \rbr{\frac{\nu}{182\mbox{~Hz}}}^{-3/2} \nonumber \\
            &\times \rbr{\frac{-\dot\nu}{7.3\E{-16}}}^{1/2} 
                    \mbox{G cm}^3,
    \end{align}
    where $\alpha$ is the misalignment angle between the rotational
    and magnetic poles. Considering $\alpha=0-90\arcdeg$, we then find a
    magnetic field strength of $B\simeq(4-6)\E8$~G at the stellar magnetic poles. 
    This magnetic field strength estimate is in line with those
    obtained for other accreting millisecond pulsars \citep[see][and
    references therein]{Mukherjee2015}.

\subsection{Orbit evolution}
\label{sec:orb}
    The observed long term binary evolution of \src is consistent with
    this source having a constant orbital period and a lower
    limit on the evolutionary timescale of
    \begin{equation}
        \tau_b = \frac{P_b}{|\dot P_b|} > 140 \mbox{~Myr}.
    \end{equation}
    Binary evolution theory predicts that systems of this type 
    evolve due to angular momentum loss through gravitational 
    radiation \citep{Kraft1962,Rappaport1982,Verbunt1993}. For
    conservative mass transfer, the binary period derivative
    is given by \citep{diSalvo2008},
    \begin{align}
        \dot P_b 
         &= -4.4\E{-11} 
          \frac{n - 1/3}{n + 5/3 - 2q}
          \rbr{\frac{P_b}{\mbox{hr}}}^{-5/3}
\nonumber \\
         &\times \rbr{\frac{M_{NS}}{\msol}}
          \rbr{\frac{M_C}{\msol}} 
          \rbr{\frac{M_{NS} + M_C}{\msol}}^{-1/3} \mbox{s~s}^{-1},
    \end{align}
    where $M_{NS}$ is the neutron star mass, $q=M_C / M_{NS}$ the binary mass ratio,
    and $-1/3 < n < 1$ the mass-radius index of the companion 
    star.
    Depending on the source inclination, \citet{Krimm2007b}
    derived a companion mass of $M_C = 0.007-0.022\msol$ for a 
    neutron star mass of $1.4\msol$. For a neutron star mass of 
    $2.2\msol$, the allowed range increased to $M_C =
    0.009-0.030\msol$. In both cases, they assumed an upper limit on
    the inclination of $i<85\arcdeg$, motivated by the fact that
    \src does not show eclipses in its light curve. Accounting
    for the extreme cases of stellar masses and $n$, 
    the binary may either be contracting or expanding. In
    either case, however, the rate of change is limited to
    $|\dot P_b| \lesssim 7\E{-14}$ s/s, which is well below
    the upper limit obtained in this work. 

    Although the binary evolution timescale we obtain for \src 
    is consistent with theory, it is worth
    noting that this is not generally true for low-mass X-ray
    binaries \citep[see][for a comprehensive discussion]{Patruno2017b}. The AMXP SAX J1808.4--3658,
    in particular, has been found to evolve on a much shorter
    timescale, with a first derivative on the orbital period of 
    $3.5\E{-12}$~s~s$^{-1}$ \citep{Hartman2008,Patruno2012a,
    Sanna2017c}. Two models have been proposed to explain this
    discrepancy: highly non-conservative mass transfer due to
    irradiation of the companion star by the pulsar
    \citep{diSalvo2008, Burderi2009}, and spin-orbit coupling in the
    companion star \citep{Hartman2008,Hartman2009}. While the latter
    depends on the companion star, and may vary from source to source,
    the former should operate in all AMXPs \citep[see also][]{
    Patruno2017a, Sanna2017b}, including \src.  The spin-down
    luminosity impinging on the companion star can be 
    estimated as 
    \begin{equation}
        \dot E_{\rm abl} = - \frac{1}{4} \rbr{\frac{R_{L2}}{a}}^2 \dot
        E_{\rm sd},
    \end{equation}
    where $\dot E_{\rm abl}$ is the ablation luminosity, 
    $R_{L2}$ is the Roche lobe radius of the companion
    \citep{Eggleton1983}, and $a$ the binary separation.
    The irradiation fraction is $f = \dot E_{\rm abl} / \dot
    E_{\rm sd}$, which, accounting for the range of allowed neutron
    star and companion masses, evaluates to $f=0.15\% - 0.35\%$. 
    The associated mass loss for the companion is given by
    \begin{equation}
        \dot M_C = -\eta \dot E_{\rm abl} \frac{R_{L2}}{G M_C},
    \end{equation}
    such that, assuming an efficiency of $\eta=100\%$,
    $\dot M_C \sim -3\E{-10}~\msol$~yr$^{-1}$. 
    The effect of this mass loss on the orbital period follows
    through the relation \citep{Frank2002}  
    \begin{equation}
        \frac{\dot P_b}{P_b} = -2\frac{\dot M_C}{M_C},
    \end{equation}
    giving a period derivative due to mass-loss of 
    $\dot P_{b, {\rm ML}} = 5\E{-12}$~s~s$^{-1}$. This value is
    well above our limit on the period derivative. Hence, in
    order for this mechanism to be consistent with our observations
    of \src, the efficiency at which the companion star converts the
    incident luminosity into mass loss must be $\eta<15\%$. This value
    is very different than the $40\%$ required in SAX~J1808.4--3658
    \citep{Patruno2016a} and is instead in line with the $<5\%$
    efficiency determined for IGR~J00291+5934 \citep{Patruno2017a}.

    ~\\

\acknowledgments
This work was supported by NASA through the \nicer mission and the
Astrophysics Explorers Program, and made use of data and software 
provided by the High Energy Astrophysics Science Archive Research Center 
(HEASARC).
P.B. was supported by an NPP fellowship at NASA Goddard Space Flight Center.  
D.A. acknowledges support from the Royal Society. 

\facilities{ADS, HEASARC, NICER}
\software{heasoft (v6.24), nicerdas (v2018-04-06 V004), tempo2
\citep{Hobbs2006}}

\bibliographystyle{fancyapj}

\end{document}